\documentstyle[12pt,epsfig]{article}
\newcommand{\beq}[1]{\begin{equation}\label{#1}}
\newcommand{\eeq}{\end{equation}}
\newcommand{\beqar}[1]{\begin{eqnarray}\label{#1}}
\newcommand{\eeqar}{\end{eqnarray}}

\newcommand{\nn}{\nonumber}
\newcommand{\D}{\partial}
\newcommand{\aD}{\stackrel{\leftrightarrow}{\partial}}
\newcommand{\el}{{\cal L}}
\newcommand{\A}{{\cal A}}
\newcommand{\tA}{\tilde{\A}}

\newcommand{\ga}{\gamma}
\newcommand{\de}{\delta}

\newcommand{\ka}{\kappa}

\newcommand{\PR}{Phys. Rev.\ }

\newcommand{\PL}{Phys. Lett.\ }
 \newcommand{\NP}{Nucl. Phys.\ }

\begin{document}
\vspace*{-2cm}
\hfill NTZ 43/97 
\vspace{4cm}
\begin{center}
{\LARGE \bf Next-to-leading gluonic reggeons  in the high-energy 
effective action\footnote{Supported by Deutsche
Forschungsgemeinschaft KI 623/1  \\ and the German-Polish agreement on
scientific and technological cooperation  N-115-95}}\\[2mm]
\vspace{1cm}
R.~Kirschner$^\dagger$ and  L.~Szymanowski$^{\dagger \#}$
\vspace{1cm}

$^\dagger$Naturwissenschaftlich-Theoretisches Zentrum  \\
und Institut f\"ur Theoretische Physik, Universit\"at Leipzig
\\ 
Augustusplatz 10, D-04109 Leipzig, Germany
\\ \vspace{2em}
$^{\#}$
Soltan Institut for Nuclear Studies, Ho\.za 69, 00-681 Warsaw, Poland
\end{center}

\vspace{1cm}
\noindent{\bf Abstract:}
We study the next-to-leading gluon exchange in the high-energy
scattering that contributes to the amplitude to order $s^0$ up to
logarithmic corrections. Similar to the leading gluon exchange
these contribution can be described in terms of reggeon
exchanges. There are several gluonic reggeons at the next-to-leading
level. Some of them transfer parity or gauge group representation 
different from the leading gluonic reggeon. Unlike the leading one 
they are sensitive to the helicity and transverse momenta of the
 scattering partons.

We extend the high-energy effective action and derive from 
the action of gluodynamics the terms describing the next-to-leading 
reggeons and their interaction in the multi-Regge approximation.

\vspace*{\fill}
\eject
\newpage

\section{Introduction}
\setcounter{equation}{0}

The leading contribution to the high-energy asymptotics of scattering
processes in QCD can be described by a reggeon and its interaction.
The exchange of two such leading reggeized gluons with interactions
summed up in the leading $\ln s$ approximation results in the BFKL
pomeron \cite{BFKL}, which is by now successfully applied in the
phenomenological analysis of semi-hard processes and in particular
of deep-inelastic scattering at small values of the Bjorken variable $x$.
The systematic improvement of the leading $\ln s$ approximation can be
organized using the reggeized gluon concept: The exchange of an even
number of leading reggeized gluons with interactions taken in the same
approximation as in the BFKL pomeron gives rise to unitarity correction
to the latter. The exchange of three reggeized gluons leads to the 
odderon \cite{BKP},
the actual role of which in phenomenology is still unclear. Unitarity
corrections to the odderon result from the exchange of an odd number of
reggeized gluons. Further next-to-leading $\ln s$ corrections
result in corrections to the scattering and production vertices and in new
vertices. These corrections are related to going beyond multi-Regge
approximation (to be explained below, sec. 2.1).

Much effort has been applied in the last years to calculate the latter
corrections \cite{beyond-multi-regge} and also to calculate the Regge
singularity induced by multiple exchange of leading reggeized gluons
\cite{mult-exch} as well as by higher reggeon interaction vertices
\cite{higher-ver}.
The reggeon concept is a starting point of the multiparticle unitarity
approach to high-energy scattering \cite{White}.

The paper is devoted to the study of contribution from gluon exchange
suppressed by one power of the cms energy squared $s$ compared to the
leading gluon exchange. We propose to apply the reggeon concept also to the
non-leading exchanges.

Analysing the high-energy asymptotics it is convenient to consider
the Mellin transform of the amplitude with respect to the energy
squared $s$. It is essentially the $t$-channel partial wave and the Mellin
variable $j$ is the complex angular momentum. The leading gluon exchange
induces Regge singularities near $j=1$, the non-leading gluon exchanges
studied here appear as reggeons with poles near $j=0$.

There are observables in high-energy scattering to which the non-leading
exchanges contribute not just as a small correction. Non-vacuum quantum
numbers like odd ($C$ or $P$) parity can be transferred by gluons.

Consider the scattering with small momentum transfer of a high-mo\-men\-tum
gluon or quark on a source of colour fields. The leading interaction
contributes to the amplitude proportional to the first power of the large
momentum. It conserves helicity of the high-momentum gluon or quark and it is
not sensitive to the details of the colour source like its distribution
in the transverse (impact parameter) plane or to its spin structure.
However such details are resolved by interactions suppressed by one power
of the large momentum compared to the leading one.

The exchange of one leading and one non-leading gluonic reggeon gives
a contribution to the small-$x$ asymptotics of the spin structure function
$g_1(x)$ of the proton \cite{BER} measuring the helicity asymmetry. 
The exchange of two non-leading gluonic
reggeons contributes to the small-$x$ asymptotics of the spin structure
function $F^{\ga}_3(x)$ of the photon (spin - 1 target), measuring the
transverse polarization asymmetry of gluons  (gluon transversity)
 \cite{EKS}.

The high-energy effective action provides a technical framework for
formulating and analysing the problems of high-energy scattering in
gauge theories. It has been proposed originally as a summary of the leading
$\ln s$ results for the leading gluon exchange in a simple form and a
starting point for going beyond this approximation \cite{Lev91}.
 Then it has been
understood that it is indeed an effective action in the sense of Wilson. 
It can
be derived from the original action by separating the fields into modes
and integrating over those modes that do not correspond neither to
scattering quanta (partons) not to exchanged quanta \cite{KLS}.
The effective action has been studied up to now in the multi-Regge approximation
for describing the leading gluon and leading fermion exchanges.
There are results going beyond this approximation \cite{Lev95}.

Here we are going to extend the procedure to non-leading gluon exchanges
contributing to ${\cal O}(s^0)$ to the amplitude. We restrict ourselves
for simplicity to the case without fermions, i.e. to pure gluodynamics. We
stay within the multi-Regge approximation improving the known procedure
by keeping terms suppressed by one power of $s$. The experience from
the exercise in high-energy scattering in (linearized) gravity
\cite{KSgrav} helped us to optimize the extensive calculations.

The structure of the effective vertices with fermions included can be
obtained from the gluonic vertices by supersymmetry transformations
relying on the similarity of QCD to supersymmetric Yang-Mills theory.
A short description of our results including a discussion about fermions
has been published earlier \cite{gr}.

Some of the technical steps in our procedure can be justified only in the
framework of perturbation theory, the applicability of which is restricted
  to the semi-hard region. There we have besides of the energy
squared a second large momentum scale (momentum transfer or virtuality
$Q^2$) which is much smaller than $s$ but still large compared to the hadronic scale.
Referring to dominating momentum configurations in this perturbative
Regge region, we can give the inverse derivatives appearing in the
calculations and in the final result a meaning, since the typical
longitudinal momenta are not small and small transverse momenta should not
be essential either. Non-local interactions are an essential feature of our
effective action.

In the next section we discuss the separation of modes according to the
multi-Regge kinematics. This is the first essential step towards the
effective action. We start from the Yang-Mills action in the light-like
axial gauge with the redundant field components eliminated.
In this way we have a direct correspondence between the fields and the
physical degrees of freedom. This gauge is convenient but can be avoided.
The resulting action does not depend on gauge. We study the impact of
the separation of modes on the triple and quartic interaction terms.
Thinking of the physical situations of high-energy gluon scattering in an
external field and of gluon-gluon quasi-elastic scattering gives us a
guideline to collect the most important terms for deriving the vertices
of the effective action.
We study the quartic terms contributing to high-energy elastic scattering.
The ones corresponding to $s$-channel gluon exchange far off-shell are
calculated in section 3. Eliminating these "heavy modes" is the second
essential step towards the effective action.

The resulting terms describing effectively quasi-elastic scattering are
analyzed in section 4. We obtain a sum of terms of the form current times
current, where the currents describe the gluons scattering with relatively
small momentum transfer. This factorizability is essential for
identifying the Regge exchanges. At the end of section 4 we write down the
effective action of quasi-elastic scattering introducing pairs of
pre-reggeon fields.

It becomes clear that due to parity symmetry, interchanging the incoming
particles, only a part of quartic terms is sufficient to obtain the
effective action for quasi-elastic scattering. This observation is used
to cut short the calculations in section 5 where we study the terms of order
5 describing effectively the inelastic $2 \to 3$ process in the multi-Regge
kinematics. By factorization we obtain the effective production vertices.
The high-energy effective action is obtained by adding the production
vertices to the effective action of quasi-elastic scattering.
The features of this action are
discussed in the last section.

\section{Separation of modes}
\setcounter{equation}{0}

\subsection{Multi-Regge kinematics}

 It is convenient to start from the
Yang-Mills action in the light-cone axial gauge $A_- = 0$,
\beqar{2.1}
    \el & = & \el^{(2)} + \el^{(3)} + \el^{(4)}  \nn \\
   \el^{(2)} & = & -2 A^{a *}(\D_+\D_- - \D\D^*) A^a \nn \\
  \el^{(3)} & = & - \frac{g}{2} J_-^a \A_+^a - \frac{g}{2} j^a \A'^a \nn \\
   \el^{(4)} & = & \frac{g^2}{8} J_-^a \D_-^{-2} J_-^a - \frac{g^2}{8} j^a j^a
\eeqar
We use light-cone components for the longitudinal part of vectors and
complex numbers for the transverse part \cite{KLS}. 
The space-time derivatives are
normalized such that $\D_+ x_- = \D_- x_+ = \D x = \D^* x^* = 1$. The gluon
field is represented by the transverse gauge potential $A^a$, $A^{a*}$.
It enters the interaction terms (\ref{2.1}) in the combinations
\beq{2.2}
\A_+^a = \D_-^{-1} (\D A^a + \D^* A^{a*})\;,\;\; \A'^a = i(\D A^a - \D^* A^{a*})
\eeq 
and in the currents
\beq{2.3}
J_-^a = i(A^* T^a\stackrel{\leftrightarrow}{\D}_- A)\;,\;\;\;\; j^a = 
(A^* T^a A) \;\;.
\eeq 
In the following we encounter besides of 
the longitudinal components $J_-^a$ also the transverse components
$J^a$, $J^{a*}$ of the vector current (obtained by replacing $\D_-$ by $\D^*$ and $\D$, 
respectively). We use the abbreviation
\beq{2.4} 
(AT^a B) =
-i f^{abc} A^b B^c 
\eeq
with $f^{abc}$ the structure constants of $SU(N)$.

The notations are chosen such that there is a close relation to the leading
terms of the effective action \cite{KLS}: The expression $\A_+^a$ 
(\ref{2.2}) describes the
leading gluonic reggeon and the current $J_-^a$ determines the leading
scattering vertex. We shall see that some of the non-leading gluonic 
reggeons are described  by the expression $\A'^a$ 
(\ref{2.2}) in terms of the original
gluon field $A^a$. 
From the point of view of momentum representation $\A_+^a$ and $\A'^a$ 
represent
the projections of the transverse gauge potential $A^a(k)$, the first parallel
to the transverse part of its momentum $\ka^\mu$ 
and the second orthogonal to $\ka^\mu$.
The non-leading scattering vertices involve besides
of the current $j^a$ other currents like $J^a$, $J^{a*}$.
Removing the redundant field components in the light-cone gauge is convenient
because now the complex field $A^a$ is directly related to the gluonic degrees
of freedom and as we have stated already in sec.~1
 introducing this gauge is a technical step which can
be avoided since the effective action is gauge invariant.

We separate the field into modes $A = A_t + A_s + A_1$.
$A_t$ are the momentum modes typical for exchanged gluons, 
$A_s$ are the modes typical for scattering gluons and $A_1$ are the heavy 
modes, which do not contribute directly to the scattering or exchange
and will be integrated out.

The modes are separated according to the multi-Regge kinematics, i.e.
the momentum configuration of a multi-particle $s$-channel (intermediate)
state ($p_l$, $l=0,1,...,n$) giving the dominant contribution in the
leading $\ln s$ approximation, see Fig. 1. Decomposing the transferred momenta
$k_l = p_A - \sum_{i=0}^{l-1}p_i$ with respect to the (almost light-like)
momenta of incoming particles $p_A$, $p_B$
\beq{2.5}
k^\mu = \sqrt{\frac{1}{s}}\left(k_+ p_B^\mu + k_- p_A^\mu\right) 
+ \ka^\mu \;\;,
\eeq
the multi-Regge kinematics is characterized by the conditions
\beqar{2.6}
&&|k_{+n}| \gg .... \gg |k_{+1}|\;\;,\;\;\; 
|k_{-n}| \ll .... \ll |k_{-1}| \nonumber
\\
&&|k_{+l}k_{-l}| \ll |\ka_l|^2 \;\;,\;\;\;\; s_l = 
|k_{-\;\,l-1}k_{+\;\,l+1}| \gg
|\ka_l|^2 \nonumber \\
&&\prod_{l=1}^n s_l = s\prod_{l=2}^n |\ka_l - \ka_{l-1}|^2 \;\;.
\eeqar
Here $\ka$ denotes the transverse (with respect to $p_A$, $p_B$) part of
the momentum $k$. It is represented by a 4-vector in (\ref{2.5}) and in the
following it will be represented by a complex number keeping the same
notation. The longitudinal momenta are strongly ordered.
The subenergies $s_l$ are large compared to the transfered momenta.
The longitudinal contribution to the transferred momenta squared is small.
In loops the main contribution from $s$-channel intermediate particles
arises from the vicinity of the mass shell. Therefore the modes
$A_t$, $A_s$, $A_1$ are characterized by the following conditions
\beqar{2.7}
&A_t&\;: \;\;|k_-k_+| \ll |\ka|^2  \nonumber \\
&A_s&\;: \;\;||k_-k_+| - |\ka|^2| \ll |\ka|^2 \nonumber \\
&A_1&\;: \;\;|k_-k_+| \gg |\ka|^2 \;\;.
\eeqar
We introduce the mode separation into the action (\ref{2.1}) by substituting
$A$ by $A_s + A_t + A_1$. The kinetic term decomposes into three, one for
each of the modes, which follows immediately from momentum conservation
\beqar{2.8}
\el_{kin}^{(0)} &=& -2A^{*a}_s (\D_+\D_- - \D\D^*)A^a_s \nonumber \\ 
&+& 2A^{*a}_t\D\D^*(1- 
\frac{\D_+\D_-}{\D\D^*})A^a_t - 2 A^{*a}_1\D_+\D_-(1- 
\frac{\D\D^*}{\D_+\D_-})A^a_1 \;\;.
\eeqar
In the kinetic term for $A_t$ and $A_1$ the second operator in the brackets
will be treated as a small one. 
In the calculations we have to keep the first order in
these corrections.

\subsection{The triple interaction}

Consider now the triple terms $\el^{(3)}$ of the action (\ref{2.1}).
We introduce the mode decomposition in  $\el^{(3)}$ and obtain many terms.
The most important terms for our discussion are those, where two of
the fields have longitudinal momenta of the same order and third one
carries much larger or much smaller longitudinal momentum. We denote by
$\el^{(3)}_1$  these terms with one of the three fields in the modes
$A_t + A_s = \tilde{A}_t$ and two in the modes $A_s + A_1 = \tilde{A}$
\beqar{2.9}
\el^{(3)}_1  &=& -\frac{g}{2}\left\{ i(\tilde{A}^*T^a\aD_- \tilde{A})
\tA^a_{+t} + (\tilde{A}^*T^a \tilde{A})\tA'^a_t + 
i(\tilde{A}^*_tT^a\aD_- \tilde{A})\tA^a_{+} \right. \nonumber \\
 &+&\left. i(\tilde{A}^*T^a\aD_- \tilde{A}_t)\tA^a_{+} + 
(\tilde{A}^*_tT^a \tilde{A})\tA'^a + 
(\tilde{A}^*T^a \tilde{A}_t)\tA'^a \right\} \;\;.
\eeqar
 $\tilde{\A}_{+t}$, $\tilde{\A}'_t$ and 
 $\tilde{\A}'$, $\tilde{\A}_+$ are
given by the expressions (\ref{2.2}) with the fields restricted to the
modes $A_t + A_s = \tilde{A}_t$ and $A_s + A_1 = \tilde{A}$, respectively.
 We rearrange
the terms in (\ref{2.9}) by using the definition (\ref{2.4}) for the
bracket $(A^*T^aA)$ and by performing integration by part in order to
put the fields with the mode
$\tilde{\A}^a_t$ as the last factor in each term (Fig. 2)
\beqar{2.10}
\el^{(3)}_1  &=& -\frac{g}{2}\left\{ i(\tilde{A}^*T^a\aD_- \tilde{A})
\tA^a_{+t} + (\tilde{A}^*T^a \tilde{A})\tA'^a_t - 
i(\tilde{A}^*T^a\aD^* \tilde{A})\tilde{A}^{*a}_{t} \right. \nonumber \\
 &-&\left. i(\tilde{A}^*T^a\aD \tilde{A})\tilde{A}^a_{t} +
i(\tilde{A}^*T^a \tilde{A})(\D\tilde{A}_t^a - \D^*\tilde{A}_t^{*a}) + 
2i(\tA_+T^a \tilde{A})\D_-\tilde{A}^{*a}_t \right. \nonumber \\ 
&-& \left.
2i(\tilde{A}^*T^a\tA_+) \D_-\tilde{A}_t^a \right\} \;\;.
\eeqar
We decompose the fields in the $\tilde{A}_t$ mode in each term into the
expressions $\A^a_+$ and $\A'^a$ (\ref{2.2}), $\D\tilde{A}_t = 
\frac{1}{2}(\D_-\tilde{A}_{+t} -i \tilde{A}'_{t})$. Writing this field
always as the last factor allows in the following to omit the
subscript $(t, s, 1)$ and the tilde referring to the range of modes.

We use the currents $J^a_-$, $J^a$, $J^{*a}$, $j^a$ introduced above 
in (\ref{2.3}) and furthermore
\beq{2.11}
J^a_s = (\frac{\D}{\D_-}AT^a A)\;,\;\;\; 
J^a_2 = (\frac{\D^*}{\D_-}A^*T^a A)
\eeq
to express the two fields in the $\tilde{A}$ mode in each term and obtain
\beqar{2.12}
\el^{(3)}_1 &=& -\frac{g}{2}\left[J^a_- - \frac{1}{2}\frac{\D_-}{\D\D^*}
(\D J^a + \D^* J^{*a}) \right. \nonumber \\
&-& \left. i\frac{\D^2_-}{\D\D^*}(\D(J^a_s + J^a_2) - \D^*(J^{*a}_s +
J^{*a}_2))\right] \A^a_+ \nonumber \\
&-&\frac{g}{2}\left[ 2j^a + \frac{i}{2}\frac{1}{\D\D^*}(\D J^a - \D^* J^{*a})
  \right. \nonumber \\
&-&\left. \frac{\D_-}{\D\D^*} (\D (J^a_s + J^a_2) + \D^* (J^{*a}_s +J^{*a}_2))
\right]\A'^a \;\;.
\eeqar
The separated triple terms (\ref{2.12})
describe in particular the interaction of
a high-energy gluon (modes $A_s$ involved in the currents) with an external field
(described by the $A_t$ modes in $\A_+$ and in $\A'$).
In the case of scattering with a large momentum component $k_-$ the term
 with $J^a_-$ gives the leading contribution of order ${\cal O}(k_-)$, the 
terms with $J^a$, $J^{*a}$ and $j^a$ contribute to order  ${\cal O}(k^0_-)$
and the other terms with $J^a_s$, $J^a_2$ result in corrections of the order
${\cal O}(k^{-1}_-)$. In describing scattering with large $k_+$ the ordering 
goes in the reverse direction.

Notice, that the $J^a_s$ terms contribute to helicity flip whereas all other
vertices conserve the helicity of the scattering gluon.

\subsection{Quartic interactions and elastic scattering}

We introduce the mode separation (\ref{2.7}) into the quartic term
$\el^{(4)}$ of the action (\ref{2.1}). We pick up the terms with
two fields in the modes $\tilde{A} = A_s + A_t$ and the other two in
all modes $A$ with the additional condition that the first two have a large
longitudinal momentum $k_-$ and the latter two have small $k_-$ but large
$k_+$,
\beqar{2.13}
\el^{(4)}_{scatt} &=& \frac{g^2}{4}\tilde{J}^a_- \frac{1}{\D_-^2}J^a_-
- \frac{g^2}{4}\tilde{j}^a j^a \nonumber \\
&-& \frac{g^2}{8} \left\{ (\tilde{A}^*T^a \aD_- A)\frac{1}{\D_-^2}
(\tilde{A}^*T^a \aD_- A) + \ldots \right\} \nonumber \\
&-& \frac{g^2}{8} \left\{ (\tilde{A}^*T^a A)(\tilde{A}^*T^a A)
+ \ldots \right\}\;\;.
\eeqar
The periods $(\dots)$ stand for the three terms obtained from the explicite
ones by shifting the tilde signs to the other fields in each of the
brackets. Now we look at the derivatives acting on the fields
$\tilde{A}$ carrying large $k_-$ modes. We apply approximations like
\beq{approx}
\frac{1}{\D_-}(\D_- \tilde{A}^* T^a A) =  (\tilde{A}^* T^a A) -
 (\frac{1}{\D_-}\tilde{A}^* T^a \D_- A) + \dots \nonumber
\eeq
and keep only terms which, after changing to momentum representation,
are of the order $k_-^1$ or $k_-^0$, and obtain
\beqar{2.14}
\el^{(4)}_{scatt} &=& \frac{g^2}{4}\tilde{J}^a_- \frac{1}{\D_-^2}J^a_-
- \frac{g^2}{4}\tilde{j}^a j^a \nonumber \\
 &-& \frac{g^2}{2}  (\tilde{A}^*T^a A)(A^* T^a \tilde{A}) +
 {\cal O}(k_-^{-1}) \;\;.
 \eeqar
We would like to write also the third term as a product of a factor involving
$\tilde{A}$ only and a second factor involving $A$. We use the relation
for the generators $T^a$ of the adjoint representation of $SU(N)$
\beq{2.15}
(T^e)_{ab}(T^e)_{cd} - (T^e)_{ac} (T^e)_{bd}  =  (T^e)_{ad} (T^e)_{cb}
\eeq
and introduce the generators $D^r$ of the reducible representation
arrising as the symmetric part in the tensor product of the two
adjoint representations of $SU(N)$ in order to write
\beq{2.16}
(T^e)_{ab}(T^e)_{cd} + (T^e)_{ac} (T^e)_{bd}  =  (D^r)_{ad} (D^r)_{cb}\;\;.
\eeq
We introduce the current
\beq{2.17}
j^r_D = (A^* D^r A) \;\;.
\eeq
Using  relations (\ref{2.15}), (\ref{2.16}) and definition (\ref{2.17})
we obtain
\beq{2.18}
\el^{(4)}_{scatt} = \frac{g^2}{4}\tilde{J}^a_- \frac{1}{\D_-^2}J^a_-
- \frac{g^2}{2}\tilde{j}^a j^a - \frac{g^2}{4}\tilde{j}^r_D j^r_D \;\;.
\eeq
The separated quartic terms (\ref{2.18})
contribute to the quasi-elastic scattering of
gluons at high-energy and limited momentum transfer.

The triple terms induce further contributions to quasi-elastic scattering:
Two vertices from $\el^{(3)}_1$ (\ref{2.12})
can be contracted by $t$-channel ($A_t$)
or by $s$-channel ($A_1$) exchanges. We consider first the
contribution of $t$-channel exchange. The contribution from heavy
modes $A_1$ will be discussed in the next section.

In one of the vertices the fields in the currents describe scattering
gluons with large $k_-$. We are going to describe high-energy scattering
in the accuracy including terms ${\cal O}(s^0)$, therefore in this
vertex we can disregard the contribution of the currents $J_s$ and $J_2$
in (\ref{2.12}).
We restrict the fields in the currents to the modes $A_s$ which are
close to mass shell. Then, up to terms ${\cal O}(k_-^{-1})$, we can
substitute $\D J +\D^* J^*$ by $\D_+ J_-$,
\beqar{2.19}
\el^{(3+)}_1 &=& - \frac{g}{2} (J^a_-  -
\frac{1}{2} \frac{\D_+ \D_-}{\D \D^*}J^a_-) \A_{+}^a \nonumber \\
&-& \frac{g}{2}\left[2j^a + \frac{i}{2} \frac{1}{\D \D^*}(\D J^a -
\D^* J^{*a})\right] \A'^a \;\;.
\eeqar
These approximations do not apply to the other vertex the currents of which
describe gluons with small $k_-$. We transform the currents using the
relation
\beqar{2.20}
\D J^a +\D^* J^{*a} &=& \D_- J^a_+ + \D_+ J^a_- + 2i(\Box A^* T^a A)
  -2i (A^* T^a \Box A)\;\;, \\
  \Box &=& \D_+ \D_- - \D \D^* \;\;,\nonumber
\eeqar
which holds  on the tree level,  and obtain
\beqar{2.21}
\el^{(3)}_1 &=& \el^{(3-)}_1 + \el^{(3-)}_2  \nonumber \\
\el^{(3-)}_1 &=& \el^{(3+)}_1 + \frac{g}{4}(\frac{\D^2_-}{\D \D^*}J^a_+)
   \A^a_{+} \nonumber \\
\el^{(3-)}_2 &=& i\frac{g}{2} \frac{\D_-}{\D\D^*} \left[ \left(
(\Box A^* T^a A) - (A^* T^a \Box A) \right) \right. \nonumber \\
&+&\left. \D_- \left( \D (J^a_s + J^a_2) - \D^* (J^{*a}_s +
J^{*a}_2)\right)\right] \A^a_{+} \nonumber \\
&+& \frac{g}{2} \frac{\D_-}{\D\D^*}\left[ \D (J^a_s + J^a_2) +
\D^* (J^{*a}_s + J^{*a}_2)\right]\A'^a
   \;\;.
\eeqar

Notice that the expression for $\el^{(3-)}_1$ does not change if we
restrict the fields in the currents to the scattering modes $A_s$.

The quartic terms induced by $t$-channel exchange are
\beq{2.22}
<\el^{(3+)}_1 \el^{(3-)}_1 >_{A_t} +
<\el^{(3+)}_1 \el^{(3-)}_2 >_{A_t} \;\;.
\eeq
The contraction $< \dots >_{A_t}$ simply means substituting in view of
the kinetic term (\ref{2.8}) the product of two exchanged fields
according to
\beq{2.23}
<\A^a_{+} \A^b_{+}>_{A_t} \to - \frac{\de_{ab}}{\D_-^2} (1 + 
\frac{\D_+ \D_-}{\D \D^*})
\;,\;\;\; <\A'^a \A'^b>_{A_t} \to \de_{ab} \;\;.
\eeq

We add the first term in (\ref{2.22}) to the original quartic terms
$\el^{(4)}_{scatt}$ (\ref{2.18}) and obtain (disregarding total
derivatives)
\beqar{2.24}
<\el^{(3+)}_1 \el^{(3-)}_1 >_{A_t} &+& \el^{(4)}_{scatt} = \nonumber \\
&=& \frac{g^2}{8} J^a_- \frac{1}{\D\D^*}J^a_+ -
\frac{g^2}{16} \D_+ J^a_- \frac{1}{\D\D^*}\D_- J^a_+  \nonumber \\
&+& g^2 j^a_s j^a_s - \frac{g^2}{2}j^a j^a
- \frac{g^2}{4} j_D^r j_D^r \;\;,
\eeqar
where
\beq{2.25}
j^a_s = j^a + \frac{i}{4}\frac{1}{\D\D^*}(\D J^a - \D^* J^{*a}) \;\;.
\eeq
is the current appearing already in $\A'$ channel in expression (2.12).
We adopt the convention that the first current factor involves the fields
describing the scattering gluons with large $k_-$ and the second current
the ones with large $k_+$. The form of this piece (\ref{2.24})
coincides, up to the
modifications to be discussed, with the effective action for quasi-elastic
scattering.

For later reference we write also the second term in expression
(\ref{2.22})  explicitely
\beqar{2.26}
&& <\el^{(3+)}_1 \el^{(3-)}_2 >_{A_t} = \nonumber \\
&=& \frac{ig^2}{4}J^a_-\left(1 + \frac{1}{2}\frac{\D_+ \D_-}{\D \D^*}\right)
\frac{1}{\D \D^*}\left[ \D(J^a_s + J^a_2) - \D^* (J^{*a}_s + J^{*a}_2)
  \right. \nonumber \\
  &+&\left. \frac{1}{\D_-}\left( (\Box A^* T^a A) -
       (A^* T^a \Box A)\right)\right] \nonumber \\
       &-& \frac{g^2}{2} j_s^a \frac{\D_-}{\D \D^*}\left[ \D(J^a_s + J^a_2)
       + \D^* (J^{*a}_s + J^{*a}_2)\right] \;\;.
\eeqar

\section{Integration over heavy modes}
\setcounter{equation}{0}

The triple vertices $\el^{(3)}_1$ induce further quartic terms by contracting
two of them with an intermediate virtual gluon in the heavy mode $A_1$.
Clearly also these terms contribute to high-energy quasi-elastic scattering.
Assuming as above that two fields in the triple vertices
(\ref{2.12}) carry large $k_-$
and restricting to the accuracy ${\cal O}(k_-^0)$ we can write these terms
as
\beq{3.1}
  < \el^{(3+)}_1 \el^{(3+)}_1 >_{A_1} \;\;.
\eeq

The other terms in $\el^{(3)}_1$ (\ref{2.12}) contribute only to order
${\cal O}(k_-^{-1})$.

The relevant part of the action for obtaining these quartic terms
is the kinetic term of the $A_1$ modes and the triple terms
$\el^{(3+)}_1$ linearized in $A_1$
\beq{3.2}
-2 A_1^{*a} \D_+ \D_- \left(1 - \frac{\D \D^*}{\D_+ \D_-}\right)A_1^a +
        \el^{(3+)}_1  \;\;.
        \eeq
The result of the approximate integration over $A_1$ in the functional
integral can be expressed by the value of this action at the saddle
point given by the Lagrangian term
\beq{3.3}
2 A^{*a}_{1C} \D_+ \D_- \left(1 - \frac{\D \D^*}{\D_+ \D_-}\right)A_{1C}^a
\;\;.
\eeq
Here $A_{1C}$ is the solution of the linearized equation of motion derived
 from (\ref{3.2})
\beqar{3.4}
&&A_{1C}  = \nonumber \\
&=& \frac{ig}{2} \left( \frac{1}{\D_+} \tilde{\A}_{+t} T^a A_s \right)
\nonumber \\
&-& \frac{ig}{2} \left( \frac{\D_-}{\D_+} \tilde{\A}_{+t} T^a
\frac{1}{\D_-} A_s \right)
- \frac{ig}{2} \left( \frac{1}{\D_+^2} \tilde{\A}_{+t} T^a
(1 - \frac{\D \D^*}{\D_+ \D_-})\D_+  A_s \right) \nonumber \\
&+& \frac{ig}{2}\left\{ \left( \frac{\D \D^*}{\D_+^2} \tilde{\A}_{+t} T^a
\frac{1}{\D_-}A_s \right) +
\left( \frac{\D^*}{\D_+^2} \tilde{\A}_{+t} T^a \frac{\D}{\D_-}A_s \right)
+ \left( \frac{\D}{\D_+^2} \tilde{\A}_{+t} T^a \frac{\D^*}{\D_-}A_s \right)
\right\}  \nonumber \\
&+& \frac{g}{2}\left( \frac{1}{\D_+} \tilde{\A}'_t T^a
\frac{1}{\D_-} A_s\right) \nonumber \\
&-& \frac{ig}{2}\left( \frac{1}{\D_+} \tilde{A}^*_t T^a
\frac{\D^*}{\D_-}A_s\right)
- \frac{ig}{2}\left( \frac{1}{\D_+} \tilde{A}_t T^a
\frac{\D}{\D_-}A_s\right) \;\;.
\eeqar
The leading contribution at large $k_-$ arises only from the first term.
We neglect terms contributing to ${\cal O}(k_-^{-1})$ and use the free
equations of motion $\Box A_s = 0$ for the scattering modes.

The two fields in $A_s$ modes in the result (\ref{3.3}) can be
expressed in terms of the currents introduced above. To illustrate
how this works we pick up the terms where the leading term in (\ref{3.4})
is multiplied with the one involving $\tilde{\A}'_t$
\beqar{3.5}
&&-\frac{ig^2}{2}\left\{\left(  A_s^* T^a \frac{1}{\D_+} \tilde{\A}_{+t}\right)
\D_+ \D_- \left( \frac{1}{\D_+}\tilde{\A}'_t T^a \frac{1}{\D_-} A_s\right)
\right. \nonumber \\
&&\left. -\left( \frac{1}{\D_-}A^*_s T^a \frac{1}{\D_+}\tilde{\A}'_t \right)
\D_+ \D_- \left(\frac{1}{\D_+}\tilde{\A}_{+t} T^a A_s  \right)\right\} \;\;.
\eeqar
We take into account that in the considered part of the mode separation
$\D_- \tilde{\A}_t$ is small compared to $\D_- A_s$
but $\D_+ \tilde{\A}_t$ is
large compared to $\D_+ A_s$ and obtain
\beq{3.6}
-\frac{ig^2}{2}\left[\left( A^*_sT^a\frac{1}{\D_+} \tilde{\A}_{+t}\right)
\left( \tilde{\A}'_tT^a A_s  \right) - \left(\frac{1}{\D_-}A^*_s T^a
\frac{1}{\D_+} \tilde{\A}'_t \right)\left(\tilde{\A}_{+t}T^a \D_- A_s \right)
   \right] \;\;.
\eeq
In this approximation we transform the expression (\ref{3.6}) by partial
integration
\beq{3.7}
-\frac{ig^2}{2}\left[\left( A^*_sT^a\frac{1}{\D_+} \tilde{\A}_{+t}\right)
\left( \tilde{\A}'_t T^a A_s  \right) - \left( A^*_s T^a
 \tilde{\A}'_t \right)\left(\frac{1}{\D_+}\tilde{\A}_{+t}T^a  A_s \right)
   \right] \;\;.
\eeq
Using the relation (\ref{2.15}) for the generators we obtain finally
\beq{3.8}
\frac{ig^2}{2}\left( A^*_sT^a A_s \right)\left( \frac{1}{\D_+}\tilde{\A}_{+t}
 T^a \tilde{\A}'_t \right) = \frac{ig^2}{2} j^a
 \left( \frac{1}{\D_+}\tilde{\A}_{+t} T^a \tilde{\A}' \right) \;\;.
 \eeq
Writing the current involving the large $k_-$ scattering modes as the first
factor we can suppress all signs refering to the modes.

In this way we obtain for the quartic terms induced by the approximate
heavy mode integration
\beqar{3.9}
&&<\el^{(3+)}_1 \el^{(3+)}_1 >_{\A_1} = \nonumber \\
&&= \frac{ig^2}{8} J^a_- \left( \frac{1}{\D_+}\A_+ T^a \A_+  \right)
\nonumber \\
&& + \frac{g^2}{16} \frac{1}{\D\D^*}\left( \D J^a + \D^* J^{*a} \right)
\left\{
i\left( \frac{1}{\D_+} \A_+ T^a \Box \frac{1}{\D_+} \A_+ \right)
\right. \nonumber \\
&&\left. + i\left( \frac{\D}{\D_+} \A_+
T^a \Box \frac{1}{\D \D_+}
\A_+  \right) + \left( \frac{\D}{\D_+} \A_+ T^a \frac{1}{\D} \A' \right)
 + c.c.
\right\} \nonumber \\
&& + \frac{ig^2}{2} j^a_s \left( \frac{1}{\D_+} \A_+ T^a \A' \right)
\nonumber \\
&& + i\frac{g^2}{16}\frac{1}{\D\D^*} \left( \D J^a - \D^* J^{*a} \right)
\left\{ i\left( \frac{\D}{\D_+} \A_+ T^a \frac{1}{\D} \A'\right)
\right. \nonumber \\
&& \left.
- \left( \frac{\D}{\D_+} \A_+ T^a \Box \frac{1}{\D\D_+} \A_+   \right)
+ c.c. \right\} \nonumber \\
&& + \frac{g^2}{8}j^r_D \left\{
\left( \frac{\D}{\D_+}\A_+ D^r \Box \frac{1}{\D \D_+}\A_+  \right)
-i \left( \frac{\D}{\D_+}\A_+ D^r \frac{1}{\D}\A' \right) \right. \nonumber \\
&&\left.  + c.c.
\right\} \;\;.
\eeqar
The fact that the two fields in $A_s$ modes carrying large $k_-$ can be
factorized in terms of the currents is the first sign of factorizability
in quasi-elastic scattering
which is essential for identifying the perturbative reggeons.

\section{Quasi-elastic scattering}
\setcounter{equation}{0}
\subsection{Quartic terms}

Summarizing the results of the previous two sections we write down the
sum of quartic terms, where two of the fields are in scattering modes 
with large $k_-$ and the two others in all modes but with small $k_-$
(see Fig. 3)
\beqar{4.1}
&&\el^{(4)}_{tot} =
< \el^{(3+)}_1 (\el^{(3-)}_1 + \el^{(3-)}_2)>_{A_t} \nonumber \\
&& +
< \el^{(3+)}_1 \el^{(3+)}_1 >_{A_1} + \el^{(4)}_{scatt} \;\;.
\eeqar
To avoid misunderstanding we say that $\el^{(4)}_{tot}$ does not mean all
quartic terms but a set of terms complete in the sense that it allows to 
extract the vertices of the effective action.

We have observed that the two fields in scattering modes factorize in 
terms of the currents. We shall first study the terms, where this current 
is $J_-^a$ or $(\D J^a + \D^* J^{*a})$. The latter can be replaced by 
$\D_+ J^a_-$ (compare Eq.(\ref{2.20})). These terms correspond to the
exchange of $\A_{t+}$.

The terms with currents $j^a$ or $j^a_s$ (Eq.(\ref{2.25})) are related to 
the exchange of $\A'_{t}$. Further there are terms with $j^r_D$
which correspond to the representation symmetric in the colour
indices in the $t$-channel, the generators of which we have denoted by
$D^r$ (\ref{2.17}).

In each case we shall consider in particular the contribution to elastic
scattering obtained by restricting also the other two fields to the
scattering modes. In this way we shall arrive at the effective action
$\el_{eff,scatt}$ describing high-energy quasi-elastic scattering.

\subsection{  $\A_+$ channel}

From Eq.(\ref{4.1}) supplemented by Eqs. (2.25), (2.27) and (3.9) 
we obtain disregarding total derivatives
\beqar{4.2}
&& \el^{(4)}_{tot}|_{\A_+} = \nonumber \\
&& = \frac{g^2}{8} J^a_- \left\{
\frac{1}{\D \D^*}\left(1 + \frac{1}{2} \frac{\D_+ \D_-}{\D \D^*}\right)
[ J^a_+ + 2i \left( \D (J^a_s + J^a_2) - \D^* (J^{*a}_s + J^{*a}_2 )
\right)
\right. \nonumber \\
&& \left.  
+ 2i\frac{1}{\D_- }\left(
 (\Box A^* T^a A) - (A^* T^a \Box A)\right)] +
 \frac{1}{2}\left(\frac{\D_+}{\D\D^*}  \right)^2 J_-^a
+i\left(\frac{1}{\D_+} \A_+ T^a \A_+\right) 
\right. \nonumber \\
&& \left. - \frac{1}{2} \frac{\D_+}{\D \D^*}\left[
i\left( \frac{\D}{\D_+} \A_+ T^a \Box \frac{1}{\D \D_+} \A_+  \right)
+ i \left( \frac{1}{\D_+} \A_+ T^a \Box \frac{1}{ \D_+} \A_+  \right)
 \right. \right. \nonumber  \\
&& \left. \left. +\left( \frac{\D}{\D_+} \A_+ T^a  \frac{1}{ \D} \A'  \right) + c.c.
\right] \right\}. 
\eeqar
If we restrict all fields to the scattering modes, where we have
$\Box A_s =0$, a number of terms vanishes obviously. We shall see that the 
contribution to quasi-elastic scattering is given by the first term only
with the substitution of $J^a_+$ by
\beq{4.3}
J^a_{R+} = i\left( A_R^* T^a 
\stackrel{\leftrightarrow}{\D}_+ A_R \right)\;,\;\;\; 
A^a_R = - \frac{\D^*}{\D} A^{*a} \;\;.
\eeq
We can say that all the other terms just result in dressing the "bare"
current to become $J^a_{R+}$. Notice that for $A_s$ modes the relation 
between $A^a_R$ and $A^a$ is just the gauge transformation which leads 
from the gauge $A^a_- =0$, which is used here, to the gauge $A^a_+ =0$.
Clearly this structure of the result is just what should be expected 
from parity symmetry, which implies the symmetry under the exchange
of indices $+$ and $-$ and the gauge transformation (\ref{4.3})
\beq{parity} 
+ \leftrightarrow -\;\;,\;\;\;
A \to A_R \;\; .
\eeq 

The contribution to quasi-elastic scattering with $\A_+$ exchange
can be written as
\beq{4.4}
\el^{(4)}_{eff,scatt}|_{\A_+} = \frac{g^2}{8} J^a_- \frac{1}{\D \D^*}J^a_{R+} -
\frac{g^2}{16} (\D_+ J^a_-)\frac{1} {(\D \D^*)^2}\D_- J^a_{R+} \;\;.
\eeq
To proof this assertion it is convenient to use the following
"dressing relations", leading to the replacement of the current $J_+$ by 
the "dressed" one $J_{R+}$,
\beqar{4.5}
&&J^a_+ + i\D\D^* \left( \frac{1}{\D_+}\A_+ T^a \A_+\right) + 2i
\left( \D (J^a_s + J^a_2) - \D^* (J^{*a}_s + J^{*a}_2)\right) \nonumber \\
&&= J^a_{R+} - i\D_+\left[ (\D_- \A_+ T^a \frac{1}{\D_+}\A_+) -
(\frac{\D}{\D^*}A T^a A) - (\frac{\D^*}{\D}A^* T^a A^*)   \right]
\eeqar
\beqar{4.6}
&& \frac{i}{2} \left( \frac{\D^*}{\D_+}\A_+ T^a \frac{1}{\D^*}\A'\right)
- \frac{i}{2} \left( \frac{\D}{\D_+}\A_+ T^a \frac{1}{\D}\A'\right)
\nonumber \\
&& = \left(\frac{\D}{\D^*}A T^a A \right) + \left(\frac{\D^*}{\D}A^* T^a A^*
 \right)
\eeqar
\beqar{4.7}
&&i\left(\D_+ J^a_- + \D_- J^a_+ \right) - 2\D_- \left( 
\D (J^a_s + J^a_2 ) - \D^* (J^{*a}_s + J^{*a}_2)  \right) \nonumber \\ 
&&+ 2\D \D^*\left(\D_- \A_+ T^a \frac{1}{\D_+} \A_+  \right) 
 = i\left(\D_+ J^a_{R-} + \D_- J^a_{R+} \right) \nonumber \\
&& + 2\D_+ \left[\D \left(\D_- A T^a \frac{1}{\D_+}\A_+   \right)
 + \D^* \left(\D_- A^* T^a \frac{1}{\D_+}\A_+     \right)   \right]
 \;\;.
\eeqar
The relations hold only if all fields involved are
 in the $A_s$ modes (which we have not indicated explicitely to simplify
the notations). They are derived by straightforward calculations, where 
the condition $\Box A_s =0$ is used repeatedly.
With these relations applied to (\ref{4.2}) we obtain (\ref{4.4})
immediately.

\subsection{ $\A'$ channel}

From Eq.(\ref{4.1}) supplemented by Eqs. (2.25), (2.27) and (3.9) we obtain
\beqar{4.8}
&&\el^{(4)}_{tot}|_{\A'} = \nonumber \\
&& = g^2 j^a_s \left\{ j^a_s - \frac{1}{2} \frac{\D_-}{\D\D^*}
\left(\D (J^a_s + J^a_2 ) + \D^* (J^{*a}_s + J^{*a}_2)   \right)
+ \frac{i}{2}\left(\frac{1}{\D_+}\A_+ T^a \A' \right)   
\right\} \nonumber \\
&&- \frac{g^2}{2} j^a j^a  \\
&&+ \frac{g^2}{4}(j^a_s - j^a)\left\{ i\left(\frac{\D}{\D_+}\A_+ T^a
\frac{1}{\D}\A' \right) - \left(\frac{\D}{\D_+}\A_+ T^a \Box 
\frac{1}{\D\D_+}\A_+  \right) + c.c. \nonumber
\right\} \;\;.
\eeqar
If we restrict all fields to the scattering modes we obtain a simple 
expression where also the two fields with small $k_-$ can be expressed 
in currents $j^a_{Rs}$ and $j^a_{R}$. These currents
are given by the expressions
for $j^a_{s}$ and $j^a$ with substituting $A^a$ by $A^a_R$ in analogy 
to Eq.(\ref{4.3}).

As in the previous subsection it is convenient to use "dressing relations"
which hold if all fields involved are in the $A_s$ modes, with
$\Box A_s =0$,
\beqar{4.9}
&&2j^a_s -j^a + i\left(\frac{1}{\D_+}\A_+ T^a \A'\right)
- \frac{1}{2}\frac{\D_-}{\D\D^*}\left(
 \D (J^a_s + J^a_2 ) + \D^* (J^{*a}_s + J^{*a}_2)\right) \nonumber \\
&& = 2j^a_{Rs} -j^a_R
\eeqar
\beq{4.10}
j^a+ \frac{i}{2}\left[\left(\frac{\D}{\D_+}\A_+ T^a \frac{1}{\D}\A'\right)
+ c.c.  \right] = j^a_R \;\;.
\eeq
Applying these relations to (\ref{4.8}) we obtain that
quasi-elastic scattering with $\A'$
exchange is described by
\beq{4.11}
\el_{eff,scatt}|_{\A'} = g^2j^a_sj^a_{Rs} -  \frac{g^2}{2} j^aj^a_R
\eeq

\subsection{Symmetric gauge group channel}

From Eq.(\ref{4.1}) supplemented by Eqs. (2.25) and (3.9) we have
\beqar{4.12}
&&\el^{(4)}_{tot}|_D = -\frac{g^2}{4}j^r_D \left\{ j^r_D 
 - \frac{1}{2}\left[ \left(\frac{\D}{\D_+}\A_+ D^r \Box 
\frac{1}{\D\D_+}\A_+ \right)  
 \right. \right. \nonumber \\
&&\left. \left. -i \left(\frac{\D}{\D_+}\A_+ D^r \frac{1}{\D}\A'\right)
+ c.c. \right]\right\} \;\;.
\eeqar
As in the other cases we study the special cofiguration where all fields are in 
the modes $A_s$. Now we have as the "dressing relation"
\beq{4.13}
j^r_D + \frac{1}{2}\left[i \left(\frac{\D}{\D_+}\A_+ D^r \frac{1}{\D}\A'\right)
+ c.c.   \right] = j^r_{RD} 
\eeq
and obtain as the symmetric gauge-group channel contribution to
quasi-elastic scattering
\beq{4.14}
\el_{eff,scatt}|_{D} = - \frac{g^2}{4}j^r_Dj^r_{RD} \;\;.
\eeq

\subsection{Effective action for quasi-elastic scattering}

The high-energy scattering of gluons at the tree level can be effectively
described by the sum of terms (\ref{4.4}),  (\ref{4.11}),  (\ref{4.14})
up to corrections of order ${\cal O}(s^{-1})$. It is remarkable
that we have a sum of terms factorized with respect to the $t$-channel.
Parity symmetry implies that the currents describing the
scattering at large $k_-$ have their counterparts in the case of scattering at
large $k_+$ obtained from the former by replacing indices
$- \to +$ and by the gauge transformation $ A \to A_R$ (\ref{parity}).
Due to this symmetry
the result has always the bilinear form of products of currents
factorizable in the $t$-channel. It is enough to study the triple terms
$\el^{(3)}_1$ describing large $k_-$ scattering and the quartic terms of
the original action $\el^{(4)}_{scatt}$ to find out what kind of terms
arise in the effective action of high-energy quasi-elastic scattering.

Indeed we see that the quartic terms 
$<\el^{(3+)}_{1}\el^{(3-)}_{1}>_{A_t} + \el^{(4)}_{scatt}$
(\ref{2.24}) coincide with the ones in (\ref{4.4}), (\ref{4.11}) and 
(\ref{4.14}) up to the index $R$ at the second current indicating the
substitution $A \to A_R$. This dressing is the only effect of the
 contribution from the heavy mode intermediate state and of terms in 
$\el^{(3)}_{1}$ negligible in the large $k_-$ scattering. The lengthy
explicit expression for the complete quartic terms $\el^{(4)}_{tot}$
is necessary only in the derivation of the effective vertices of gluon
production discussed in the next section.

We introduce 5 pairs of pre-reggeon fields, one pair for each current
 times current term. In this way the quasi elastic high-energy scattering
can be described by the following effective action
\beqar{4.15}
&&{\el}_{eff, scatt} = {\el}_{kin} + {\el}_{s -} + {\el}_{s +} \nn \\
&&{\el}_{kin} = - 2 A_s^{a*}(\D_- \D_+ - \D \D^* )A_s^a
-2 \A_+^a \D \D^* \A_-^a + \A_{(+)}^a \A_{(-)}^a \nn \\
&& - \A'^{a(+)}_s \A'^{a(-)}_s + 2\A'^{a(+)}_2 \A'^{a(-)}_2
 + B^{r(+)}B^{r(-)}  \\
&& {\el}_{s -} = - \frac{g}{2}J_-^a \A_+^a - \frac{g}{4} (\frac{\D_+}{\D
\D^*}J_-^a) \A_{(+)}^a -g j_s^a \A_s'^{a(+)} - gj^a\A'^{a(+)}_2 -
\frac{g}{2}j_D^r B^{r(+)} \nonumber \;\;.
\eeqar
${\el}_{s +}$ is obtained from ${\el}_{s -}$ by replacing the labels $+
\leftrightarrow -$ and the currents by their partners with label $R$.
This result is checked  simply by observing that each of the 
quartic scattering terms (\ref{4.4}), (\ref{4.11}), (\ref{4.14})
is reproduced from (\ref{4.15}) by integrating out
the corresponding pre-reggeon.

Each pre-reggeon is represented by a pair of fields. The distinction by the 
labels $+$ and $-$ is related to the fact that the reggeon exchange
is oriented in rapidity: The pre-reggeon fields with label $+$ couple
to scattering gluons with large $k_-$ only and vice versa.

The pre-reggeon ($\A^a_+, \A^a_-$) is the leading one. It contributes as $s^1$
to the amplitude and it is well known that the exchange of two of them,
interacting via emission and absorption of $s$-channel gluons, results
in the BFKL pomeron. The pre-reggeon  ($\A^a_{(+)}, \A^a_{(-)}$)
can be regarded as the non-leading partner of the first one.
Both conserve helicity of the scattering gluon and both carry positive
$P$-parity and negative $C$-parity.

In the $\A'$- channel, considered in the sub-section 4.3 we encounter two
pre-reggeons at the non-leading level ${\cal O}(s^0)$, $\A_s'^{a(\pm)}$ and
$\A'^{a(\pm)}_2$. Both conserve helicity but their vertices depend on the
sign of the helicity. They carry negative $P$- and $C$- parities.
The difference between them is not obvious. At this point we just notice
that the coupling of   $\A_s'^{a(\pm)}$ involves a term sensitive to 
the transverse momenta of the scattering gluons whereas $\A'^{a}_2$ is
insensitive to transverse momenta.

The pre-reggeons discussed so far are in the gauge-group state of 
the gluon.
The $t$-channel factorization
unavoidably leads also to exchanges in other gauge-group
representations, which arise as the symmetric part in the tensor product of
two adjoint representations. Among the corresponding pre-reggeons
there is a colour singlet one.  

\section{Inelastic scattering}
\setcounter{equation}{0}

\subsection{Terms of order 5}

The essential ingredients of the high-energy effective action beyond the
effective action of quasi-elastic scattering are the production vertices.
In general these vertices can be obtained from the effective interaction
terms of order 5, where the fields are in scattering modes $A_s$ and their
longitudinal momenta correspond to the kinematics of an inelastic
$2 \to 3$ scattering in multi-Regge kinematics (\ref{2.6}), which we denote
by $\el^{(5)}$. The production vertices, triple vertices involving two
pre-reggeons and one scattering particle, are obtained by factorizing the
scattering vertices $\el_{s-}$ and $\el_{s+}$
\beq{5.1}
\el^{(5)} = < \el_{s-} \el_{p} \el_{s+}> \;\;.
\eeq
Here the brackets stand for contraction of the pre-reggeons by
substituting their product by 
 propagators read off from the kinetic terms in
$\el_{eff,scatt}$ (\ref{4.15}).

A contribution to $\el^{(5)}$ in the desired mode configuration is obtained
from $\el^{(4)}_{tot}$ by contracting with $\el^{(3-)}_1$,
$<\el^{(4)}_{tot} \el^{(3-)}_1>_{A_t}$. $\el^{(4)}_{tot}$ involves two
fields in $A_s$ mode with large $k_-$ and $\el^{(3-)}_1$ two fields in
$A_s$ mode with large $k_+$. One of the two other fields in
$\el^{(4)}_{tot}$ is to be restricted to the scattering mode and the other
to the exchange mode.

In this contribution
the factorization in the two $t$-channels is immediate, because the pair
of fields with large $k_-$ is factorized in $\el^{(4)}_{tot}$
(\ref{4.2}), (\ref{4.8}), (\ref{4.12}) and the pair of fields with
large $k_+$ is factorized in  $\el^{(3-)}_1$. Therefore the production
vertices involving only those pre-reggeons the couplings (currents)
of which appear in $\el^{(3\pm)}_1$ can be read off directly from
$\el^{(4)}_{tot}$. This does not apply only to one of the odd-parity
reggeons $\A'^{\pm}$ and to the gauge-group symmetric pre-reggeons
$B^{r(\pm)}$.

Besides of the terms discussed we have more contributions.
To order 5 the interaction 
terms in the mode configuration of interest are given by
(Fig.~4)
\beqar{5.2}
&& \el^{(5)} = <\el^{(4)}_{tot} \el^{(3-)}_1>_{A_t}
+ <\el^{(4)}_{tot} \el^{(3-)}_2>_{A_t}
+ <\el^{(4)}_{tot} \el^{(3+)}_1>_{A_1} \nonumber \\
&& + <\el^{(3+)}_{1} \el^{(4)}_{scatt}>_{A_t}
+ <\el^{(3+)}_1 \el^{(4)}_{scatt}>_{A_1} \;\;.
\eeqar

In sec.4 we have seen that the heavy mode contributions 
and the contribution with $\el^{(3-)}_2 $ merely lead to
the "dressing" of the currents involving the large $k_+$ scattering modes.
Analogously, the second and third terms
 in (\ref{5.2}) contributes only to dressing
the currents in $\el_{s+}$ (\ref{4.15}). The structure of the result
appears already in the sum of the other terms and the effective production
vertices can be obtained without writing the second and the third terms
 in (\ref{5.2})
explicitely.

\subsection{$\A_- - \A_+$ exchanges}

As discussed above the production vertices involving the even-parity
pre-reggeons $\A_\pm$ and $ \A_{(\pm)}$ can be read off directly from the
complete quartic terms $\el^{(4)}_{tot}|_{\A_+}$ (\ref{4.2}).
We recall that $J^a_-$ involves the two scattering fields with large
$k_-$. One of the other two fields has to be restricted to the
scattering modes too. We 
may restrict ourselves to  the contributions where the inelastic
gluon corresponds to the field $A^*_s$. The production vertex with $A_s$
is then obtained by complex conjugation.
The remaining field in $\el^{(4)}_{tot}$
is to be projected onto $\A_+$ in the $t$-channel
modes, because the wanted contribution
 $<~\el^{(4)}_{tot} \el^{(3-)}_1>$ to
$\el^{(5)}_p$ is obtained by contraction with \\
$\el^{(3-)}_1|_{\A_+} =
\frac{g}{4}\left(\frac{\D_-^2}{\D\D^*} J^a_+\right) \A^a_{t+}$.

For illustration we pick up in $\el^{(4)}_{tot}$ (\ref{4.2}) the term
$i\frac{g^2}{8}J^a_- \left(\frac{1}{\D_+}\A_+ T^a \A_+\right)$. Its
contribution to the inelastic production with $A^*_s$ is given by
\beq{5.3}
i\frac{g^3}{32}\left\{ \left(J_- T^a (1+\frac{\D_+\D_-}{\D\D^*})
\frac{1}{\D\D^*}J_+ \right)
+ \D_+ \left(J_- T^a (1+\frac{\D_+\D_-}{\D\D^*})\frac{1}{\D_+}
\frac{1}{\D\D^*}J_+ \right)
\right\}\frac{1}{\D}A^{*a}_s \;\;.
\eeq
For the leading exchange related to $J_\pm$ the correction to the propagator
$(1+\frac{\D_+\D_-}{\D\D^*})$
has to be included. Up to  terms contributing to order 
${\cal O}(\frac{1}{s})$ we obtain
\beqar{5.4}
&& i\frac{g^3}{16}\left\{ 2\left(J_- T^a \frac{1}{\D\D^*}J_+ \right)
+ \left(\D_+ J_- T^a \frac{\D_-}{(\D\D^*)^2}J_+ \right) 
 + 2\D_+ \left(J_- T^a \frac{\D_-}{(\D\D^*)^2}J_+ \right) \right. \nonumber \\
&&\left. + \left(\D_+ J_- T^a \frac{1}{\D\D^* \D_+}J_+ \right)
   \right\}
\frac{1}{\D}A^{*a}_s \;\;.
\eeqar
The last term can be transformed in the following way
\beqar{5.5}
&& \left( \D_+ J_- T^a \frac{1}{\D\D^* \D_+}J_+ \right) 
\frac{1}{\D}A_s^{*a} = \nonumber \\
&& = - \left(  \D_+ J_- T^a \frac{1}{\D\D^* }J_+  \right)
\frac{\D_-}{\D^2 \D^*}A_s^{*a}
+ \left(  \D_+^2 J_- T^a \frac{1}{\D\D^* }J_+  \right)
\frac{1}{\D_+^2 \D}A_s^{*a} \;.
\eeqar
The second term gives a small contribution. We have used the condition
$\Box A_s =0$.

Factorizing $\el_{s\pm}$ we obtain the corresponding contribution to
$\el_p$,
\beqar{5.6}
&& ig\left\{ \left(\D\D^* \A_- T^a \A_+  \right)
- \frac{1}{2}\left( \D\D^* \A_{(-)} T^a \frac{1}{\D\D^*} \A_{(+)} \right)
\right. \nonumber \\
&& \left. - \D_+\left( \D\D^* \A_{-} T^a \frac{1}{\D\D^*} \A_{(+)} \right)
-  \frac{1}{2} \frac{\D_-}{\D\D^*}
\left(\D\D^* \A_{(-)} T^a \A_+  \right)\right\}\frac{1}{\D}A_s^{*a}
\eeqar

Repeating the outlined procedure with all terms in $\el^{(4)}_{tot}$
(\ref{4.2}) we obtain the effective production vertices in the channels
with the pre-reggeons $\A_\pm$ and their non-leading partners
$\A_{(\pm)}$, 
\beqar{5.7}
&&\el_p|_{\A_\pm} = \nonumber \\
&&= - \frac{ig}{2}\left\{ \left[
2\left( \D^* \A_- T^a \D \A_+  \right)
+ \frac{1}{2}\left(  \A_{(-)} T^a \D \A_{(+)}  \right)
\right. \right. \nonumber \\
&& \left. \left.
+ \frac{1}{2}\left( \frac{\D^*}{\D} \A_{(-)} T^a \frac{\D}{\D^*} \A_{(+)}
\right)\right] \frac{1}{\D} A^{*a} \right. \nonumber \\
&&\left. +\left[ \left( \frac{1}{\D} \A_{(-)} T^a \D \A_+  \right)
+ \frac{1}{\D\D^*}\left( \D^* \A_{(-)} T^a \D \A_+  \right)
\right. \right. \nonumber \\
&& \left. \left.
+  \frac{1}{\D\D^*}\left(  \A_{(-)} T^a \D\D^*  \A_+  \right)
\right] \frac{\D_-}{\D} A^{*a} \right. \nonumber \\
&&\left. + \left[ \left( \D^* \A_- T^a \frac{1}{\D^*} \A_{(+)}  \right)
+ \frac{1}{\D\D^*}\left( \D^* \A_- T^a \D \A_{(+)}  \right)
\right. \right. \nonumber \\
&& \left. \left.
+  \frac{1}{\D\D^*}\left(  \D\D^* \A_- T^a   \A_{(+)}  \right)
\right] \frac{\D_+}{\D} A^{*a}\right\} + c.c. \;\;.
\eeqar

\subsection{ $\A' - \A'$ exchange}

We consider first the contribution in the term
$<\el^{(4)}_{tot}\el^{(3-)}_{1}>|_{A_t}$ to these channels obtained by
combining $\el^{(4)}_{tot}|_{\A'}$ (\ref{4.8}) with $\el^{(3-)}_{1}|_{\A'}
= -gj^a_s \A'^a$,
 (\ref{2.21}),
\beqar{5.8}
&&<\el^{(4)}_{tot}\el^{(3-)}_{1}>|_{\A'-\A'} = \nonumber \\
&& = -i\frac{g^3}{4} \left\{ 2\left(\frac{1}{\D}j_s T^a \D j_s \right)
+ \left(\D^* j_s T^a \frac{1}{\D^*}j_s \right) \right. \nonumber \\
&&\left. + \left(\frac{\D^*}{\D}j_s T^a \frac{\D}{\D^*}j_s \right)
+ \D^* \left((j_s - j) T^a \frac{1}{\D^*}j_s \right)\right\}
\frac{1}{\D} A^{*a}_s  + c.c. \;\;.
\eeqar
Now we calculate the contribution of 
$<\el^{(3+)}_{1}\el^{(4)}_{scatt}>|_{A_t}$ using
\beq{5.9}
\el^{(3+)}_{1}|_{\A'} = -g j^a_s \A'^a \;,\;\;\; \el^{(4)}_{scatt}|_{\A'}=
- \frac{g^2}{2} j^a j^a \;\;.
\eeq
One of the two fields in the first current factor in 
$\el^{(4)}_{scatt}|_{\A'}$ corresponds to the produced particle 
and the other involves the $t$-channel modes to be contracted with
$\el^{(3+)}_{1}|_{\A'}$,
\beq{5.10}
\el^{(4)}_{scatt}|_{\A'} \to \frac{ig^2}{4}\left(\frac{1}{\D}\A' T^a j \right)
A^{*a} + c.c. \;\;\;.
\eeq
This implies
\beq{5.11}
<\el^{(3+)}_{1}\el^{(4)}_{scatt}>_{A_t}|_{\A'-\A'} =
-\frac{ig^3}{4}\left(\frac{1}{\D}j_s T^a j \right)A^{*a} + c.c. \;\;\;.
\eeq
The next term to be evaluated is 
$<\el^{(4)}_{scatt}\el^{(3+)}_{1}>_{A_1}$ supplying
 the brems\-strah\-lung correction to the original quartic vertex 
caused by the leading term in $\el^{(3)}_{1}$   (2.12).
In section 3 we have done the calculation for the heavy mode contributions
 $\el^{(3)}_{1}|_{\A'}  = -g j^a \A'^a$ resulting in Eq.(\ref{3.8}).
The heavy mode contribution involving the quartic vertex
$ \el^{(4)}_{scatt}|_{\A'}$ (\ref{5.9})
 is given 
by the analogous expression with the substitution $\A'^a \to \frac{g}{2}j^a$.
This gives immediately the result:
\beq{5.12}
<\el^{(4)}_{scatt}\el^{(3+)}_{1}>_{A_1} = -\frac{ig^3}{4}\left(j T^a j\right)
\frac{1}{\D}A^{*a} + c.c. \;\;.
\eeq
For the sum of the calculated contributions we have
\beqar{5.13}
&&(<\el^{(4)}_{tot}\el^{(3-)}_{1}>_{A_t}
+ <\el^{(3+)}_{1}\el^{(4)}_{scatt.}>_{A_t}
+ <\el^{(4)}_{scatt}\el^{(3+)}_{1}>_{A_1})|_{\A'-\A'} \nonumber \\
&& = -\frac{ig^3}{4}\left\{ 2\left(\frac{1}{\D}j_s T^a \D j_s \right)
+ 2\left(\D^* j_s T^a \frac{1}{\D^*} j_s \right) \right. \nonumber \\
&& \left. + \left(\frac{\D^*}{\D} j_s T^a \frac{\D}{\D^*} j_s  \right)
+\left(j_s T^a j_s  \right) +\left(j T^a j  \right)  \right. \nonumber \\
&& \left. - \left(\frac{1}{\D}j_s T^a \D j_s \right)
- \left(j_s T^a j  \right) \right. \nonumber \\
&& \left. - \left(\D^* j T^a \frac{1}{\D^*} j_s \right)
- \left(j T^a j_s  \right)\right\} \frac{1}{\D} A^{*a} + c.c. \;\;.
\eeqar
The factorization with respect to $\el_{s\pm}$ amounts to the replacement \\
$-g j^a_s \to \A'^a_s$, $\frac{g}{2}j^a \to \A'^a$ and
we obtain the effective production vertices
\beqar{5.14}
&&\el_p|_{\A'-\A'} = \nonumber \\
&& = -\frac{ig}{4}\left\{ 
2\left(\frac{1}{\D} \A'^{(-)}_s T^a \D \A'^{(+)}_s \right)
+ 2\left(\D^* \A'^{(-)}_s T^a \frac{1}{\D^*} \A'^{(+)}_s \right) \right. \nonumber \\
&& \left. + \left(\frac{\D^*}{\D} \A'^{(-)}_s T^a \frac{\D}{\D^*} 
\A'^{(+)}_s \right)
+\left(\A'^{(-)}_s T^a \A'^{(+)}_s  \right) 
+4 \left(\A'^{(-)}_2 T^a \A'^{(+)}_2 
 \right) \right.
\nonumber \\
&&\left. +2\left(\frac{1}{\D} \A'^{(-)}_s T^a \D \A'^{(+)}_2 \right) + 2
\left(\A'^{(-)}_s T^a \A'^{(+)}_2 \right) \right. \nonumber \\
&&\left. + 2\left(\D^* \A'^{(-)}_2 T^a \frac{1}{\D^*} \A'^{(+)}_s \right)
+ 2\left(\A'^{(-)}_2 T^a \A'^{(+)}_s  \right)\right\} \frac{1}{\D}A^{*a} + c.c.\;\;.
\eeqar

\subsection{$\A'-\A_+$ exchanges}

Unlike the previous case
 the production vertices can be obtained from the
quartic terms $\el^{(4)}_{tot}|_{\A'}$ (\ref{4.8}).
We calculate the part $<\el^{(4)}_{tot}\el^{(3-)}_{1}>$ obtained by
contracting the quartic terms with $\el^{(3-)}_{1}|_{\A_+}$,
\beqar{5.15}
&&<\el^{(4)}_{tot}\el^{(3-)}_{1}>|_{\A'-\A_+} = \nonumber \\
&& = \frac{g^3}{8}\left\{ \D_- \left( \frac{1}{\D}j_s T^a \frac{1}{\D^*}J_+
 \right) + \frac{\D_-}{\D}\left( j_s T^a \frac{1}{\D^*}J_+ \right)
 - \frac{\D_-}{\D}\left( j T^a \frac{1}{\D^*}J_+ \right) \right.
  \\
 &&\left. + \frac{1}{2} \D^* \left( (j_s - j) T^a \frac{\D_-}{\D^{*2}\D}J_+ \right)
 + \frac{1}{2}\D \left( \frac{\D^*}{\D}j_s T^a \frac{\D_-}{\D^{*2}\D}J_+
 \right)\right\} \frac{1}{\D}A^{*a}_s + c.c. \;\;. \nonumber
 \eeqar
The remaining contributions in (\ref{5.2}) lead merely to the "dressing"
substitution $J_+ \to J_{R+}$. The corresponding contribution to the
effective production vertices is obtained factorizing
$\el_{s+}|_{\A'}$ and $\el_{s-}|_{\A_+}$. The result is obtained from
(\ref{5.15}) by the following substitutions:\\
$-gj^a_s \to \A'^a_s$, $\frac{g}{2}j^a \to \A'^a$, $-\frac{g}{4}
\frac{1}{\D\D^*}J^a_{R+} \to
\A^a_+$, $\frac{g}{4}\frac{\D_-}{\D\D^*}J^a_{R+} \to \A^a_{(+)}$,
\beqar{5.16}
&&\el_p|_{\A'-\A_+} = \nonumber \\
&& = -\frac{g}{2}\left\{ -\D_- \left(\frac{1}{\D}\A'^{(-)}_s T^a \D \A_+\right)
- \frac{\D_-}{\D} \left(\A'^{(-)}_s T^a \D \A_+ \right) \right. \nonumber \\
&&\left. - 2 \frac{\D_-}{\D}\left(\A'^{(-)}_2 T^a \D \A_+ \right)
+ \D^*\left(\A'^{(-)}_2 T^a \frac{1}{\D^*}\A_{(+)}   \right)\right.
 \\
&&\left. + \frac{1}{2} \D^*\left(\A'^{(-)}_s T^a \frac{1}{\D^*}\A_{(+)}\right)
+  \frac{1}{2}\D \left( \frac{\D^*}{\D}\A'^{(-)}_s T^a \frac{1}{\D^*}\A_{(+)}
\right) \right\} \frac{1}{\D}A^{*a}_s + c.c. \;\;. \nonumber
\eeqar

The case where the reggeon $\A_\pm$ couples to the incomming particle
with large $k_-$ and $\A'$ to the one with large $k_+$ looks more
complicated. Similar to the $\A'-\A'$ case in subsection 5.3
besides of $<\el^{(4)}_{tot} \el^{(3-)}_{1}>$ more contributions have to
be evaluated explicitely. The result $\el_p|_{\A_- - \A'}$ can be
obtained however from the above expression (\ref{5.16}) by interchanging
the labels $+ \leftrightarrow -$ on the reggeon field and longitudinal
derivatives and by replacing $\frac{1}{\D}A^{*a}_s \leftrightarrow
-\frac{1}{\D^*}A^a_s$, which is just the $P$-parity transformation
(\ref{parity}).

\subsection{Vertices with colour-symmetric  pre-reggeons}

Analogous as in the previous subsection we consider first the case that
the non-octet reggeon $B^{r(\pm)}$ is in the $t$-channel connected to the
large $k_-$ scattering particles. It is not difficult to check that
 there is no gluon production vertex with the reggeons $B^{r(\pm)}$
 in both $t$-channels. The vertex with $B^{r(+)}$ coupling to large $k_-$
 can be read off from the corresponding terms in
 $<\el^{(4)}_{tot} \el^{(3-)}_{1}>$. The relevant projection is (\ref{4.12}),
\beqar{5.17}
&&\el^{(4)}_{tot}|_D = \nonumber \\
&& = \frac{g^2}{8} j^r_D\left\{
\left( \frac{\D^*}{\D_+}\A_+ D^r \Box \frac{1}{\D^*\D_+}\A_+ \right)
+ i\left( \frac{\D^*}{\D_+}\A_+ D^r  \frac{1}{\D^*}\A' \right) + c.c.
\right\} \nonumber \\
&& - \frac{g^2}{4} j^r_D j^r_D \;\;.
\eeqar
We consider first the case of the reggeons $\A_\pm$ or $\A_{(\pm)}$ in the
other $t$-channel, i.e. we contract
\beq{5.18}
\el^{(3-)}_{1}|_{\A_+} = \frac{g}{4} 
\left(\frac{\D^2_-}{\D\D^*} J^a_+\right) \A^a_+
\eeq
and obtain
\beqar{5.19}
&&<\el^{(4)}_{tot}\el^{(3-)}_{1}>_{B^{(-)}-\A_+} = \nonumber \\
&& = \frac{g^3}{32} j^r_D \left\{
2\left(\frac{\D^*}{\D}A^*_s D^r \frac{1}{\D^*\D_+}J_+ \right)
+ \left(\frac{\D^*}{\D}A^*_s D^r \frac{\D_-}{\D\D^{*2}}J_+ \right)\right\}
\nonumber \\ 
&& + c.c. \;\;.
\eeqar
We recall the notation for the bracket with $D^r$
\beq{5.20}
\left(\A_1 D^r \A_2 \right) B^r = \A^a_1 \A^b_2 B^r D^r_{ab}
\eeq
and introduce the notation
\beq{5.21}
\left(B \tilde{D}^a \A_2\right)\A^a_1 = \A^a_1 \A^b_2 B^r D^r_{ab}
\eeq
for the same expression for convenience in order to write
\beqar{5.22}
&&<\el^{(4)}_{tot}\el^{(3-)}_{1}>_{B^{(-)}-\A_+} =  \\
&& = \frac{g^3}{32}\left\{
-2\frac{\D_-}{\D}\left(j_D\tilde{D}^a \frac{1}{\D^*}J_+ \right) -
\D^*\left(j_D\tilde{D}^a \frac{\D_-}{\D \D^{*2}}J_+ \right)\right\}
\frac{1}{\D}A^{*a}_s \nonumber \\
&&+ c.c. \;\;.
\eeqar
The remaining contributions in (\ref{5.2}) lead to the dressing substitution
$J_+ \to J_{R+}$. We obtain the effective production vertices by
factorising $\el_{s\pm}$,
\beqar{5.23}
&&\el_p|_{B^{(-)}- \A_+} = \nonumber \\
&& = \frac{g}{2}\left\{ \frac{\D_-}{\D}\left( B^{(-)}\tilde{D}^a \D \A_+
\right) 
 - \frac{1}{2}\D^*\left( B^{(-)}\tilde{D}^a \frac{1}{\D^*} 
\A_{(+)} \right)
\right\} \frac{1}{\D}A^{a*} \nonumber \\
&&+ c.c. \;\;.
\eeqar

The vertex $\el_p|_{\A_- -B^{(+)}}$ can be obtained by the substitutions
discussed at the end of section 5.4 which are implied by parity symmetry.
The calculation directly from
(\ref{5.2}) takes to evaluate more terms.

Now we consider the case of a pre-reggeon $\A'_s$ in the second $t$-channel.
There is no vertex with $B^{r(-)}$ in one and $\A'^{(+)}_2$ in the other
$t$-channel. This is easily seen because $\el^{(3-)}_1$ does involve $j_s$
but not $j$.

We contract $\el^{(4)}_{tot}|_D$ with
\beq{5.24}
\el^{(3-)}_{1}|_{\A'} = -g j^a_s \A'^a
\eeq
and obtain
\beqar{5.25}
&&<\el^{(4)}_{tot}\el^{(3-)}_{1}>|_{B-\A' } = \nonumber \\
&& = \frac{ig^3}{8}\D^* \left(j_D \tilde{D}^a \frac{1}{\D^*} j_s \right)
\frac{1}{\D}A^{*a} + c.c. \;\;.
\eeqar

The remaining contribution in (\ref{5.2}) lead to the dressing substitution \\
$j_s \to j_{Rs} $. We obtain the effective production vertex by factorizing
$\el_{s\pm}$,
\beq{5.26}
\el_p|_{B-\A'} = -\frac{ig}{4} \D^* \left(B^{(-)} \tilde{D}^a  \frac{1}{\D^*}
 A'^{(+)}_s   \right) \frac{1}{\D}A^{*a} + c.c. \;\;.
\eeq

The vertex $\el_p|_{\A' - B}$ can be obtained by the parity substitution
(\ref{parity}).

\section{The high-energy effective action}
\setcounter{equation}{0}

The effective action of high-energy scattering is obtained from the 
effective action of quasi-elastic scattering (\ref{4.15}) by adding the 
effective production vertices $\el_p$
\beq{6.1}
\el_{eff} = \el_{eff, scatt} + \el_p \;\;.
\eeq
The production vertices $\el_p$ have been obtained for all pre-reggeons
 appearing in gluodynamics in the vicinity of $j=1$ and $j=0$ in
Eqs.(\ref{5.7}), (\ref{5.14}), (\ref{5.16}), (\ref{5.26})
\beqar{6.2}
\el_p &=& \el_p|_{\A_- - \A_+ } + 
\el_p|_{\A' - \A'} + \el_p|_{\A' - \A_+} + \el_p|_{\A_- - \A'} \nonumber \\
&+& \el_p|_{B - \A_+} + \el_p|_{\A_- - B} + \el_p|_{\A' - B} 
 + \el_p|_{B - \A'} \;\;.
\eeqar

The effective action can be written in terms of the complex field \\
$\phi^a = i\frac{1}{\D}A^{*a}_s$, describing the two helicity states of
scattering gluons, and the pre-reggeon fields $\A_\pm$,
$\A_{(\pm)}$, $\A'^{(\pm)}_s$,  $\A'^{(\pm)}$, $B^{(\pm)}$.

The parity transformation, interchanging the incoming particles, acts on the
terms of the action by interchanging labels $+ \leftrightarrow -$ 
accompanied by interchanging $\phi$ and $\phi^*$ (i.e. inversion of the
helicities of  scattering gluons).
$\el_{kin}$ is symmetric under this transformation. The currents
representing the scattering vertices $J_-$, $J$, $J^*$, $j_s$, $j$, $j_D$
of large $k_-$ transform into the corresponding currents of vertices with
large $k_+$, $J_{R+}$, $J_R$, $J^*_R$, $j_{Rs}$, $j_R$, $j_{RD}$, 
respectively. This means that $\el_{s+}$ and  $\el_{s-}$ transform into
each other.
The production vertices $\el_p$ are symmetric under the parity transformation.
In this way we check that the whole effective action is parity symmetric,
as expected.

In the effective action the longitudinal and transverse space-time
dimensions are separated to a large extend. Besides of the kinetic
term of the scattering fields $\phi$, all terms are invariant under
separate scale transformations in longitudinal and in transverse directions.
All terms are invariant with respect to longitudinal 
Lorentz boosts and with respect to
rotations in the transverse plane.

The notations for the pre-reggeon fields have been choosen 
in such a way that they all
 are invariant with respect to transverse rotations. The rotation
acts on the complex numbers $x$ describing the impact parameters
as
\beq{6.3}
x \to e^{i\alpha} x \;\;.
\eeq
The (transverse) gluon fields $A$ transform in the same way and the 
derivatives in the opposite way: $\D \to  e^{-i\alpha} \D$.

We have written all vertices in $\el_p$ in such a way that no transverse
derivatives act on $\phi^a = i\frac{1}{\D}A^{*a}_s$ or $\phi^*$,
which are invariant under (\ref{6.3}). We consider the vertices  as
transitions of a "left" (coupling to large $k_-$ gluons) to a "right"
(coupling to large $k_+$ gluons) pre-reggeon. Consider for example
the leading pre-reggeon vertices
\beq{6.4}
-ig\left[\left(\D^* \A_- T^a \D \A_+   \right)\phi^{*a}
+ \left(\D \A_- T^a \D^* \A_+   \right)\phi^{a}  \right] \;\;.
\eeq

In the production term of $\phi^*$
the state related to the left reggeon $\A_-$ transforms as $e^{i\alpha}$,
i.e. it has transverse angular momentum $n=1$.
One can read this term as a transfer of the angular momentum $n=1$
from the "left" to the "right" pre-reggeon.
 In the production vertex of $\phi$ the angular momentum $n=-1$
is transferred from the left to the right.

We observe that in general non-negative angular momentum $n \ge 0$ is
transferred from left to right in all production vertices of $\phi^*$
whereas in the terms of $\phi$ we have the transfer of $n \leq 0$ only.

In the $\phi^*$ vertices we encounter $n=1$ in the leading $(\A_-, \A_+)$
vertex, 
$n=0,1,2$ in the other vertices.

Notice, in particular, that the transfer of $n=1$ is absent in the
non-leading positive-parity pre-reggeons $\A_{(\pm)}$ and that there
is only zero angular momentum transfer in the ($\A'^{(-)}_2,  \A'^{(+)}_2$)
 vertex. 

\section{Summary}

In this paper we have extended the high-energy effective action
to the next-to-leading exchange contributing to ${\cal O}(s^0)$
to the amplitude. Staying within the multi-Regge approximation,
i.e.  assuming the $s$-channel multiparticle intermediate states
to obey the conditions (\ref{2.6}), we improve the approximation in all
transformations related to  the separation of modes, integration over
heavy modes and extraction of effective vertices keeping all terms
suppressed by one power of $s$ compared to the leading ones.
We have introduced  pre-reggeon fields describing in the effective
action the next-to-leading exchanges.
For this the factorizability of the quartic terms for high-energy
scattering is essential. We encounter several pre-reggeons
at the level  ${\cal O}(s^0)$: Some of them carry the gauge group 
representation of the gluon and some of them other representations.
Among the first ones we have the next-to-leading partner of the 
leading reggeized gluon carrying positive $P$-parity and insensitive
to the helicity of scattering partons. Two pre-reggeons carry odd
$P$-parity and are sensitive to the helicity.
The physical difference between the latter is one of the questions
to be investigated in more detail. One of them is sensitive to the 
transverse momenta and the other not. More general, it has to be 
investigated whether all these pre-reggeons reggeize in the same 
way as the leading gluon exchange.

The effective vertices describing emission or absorption of a gluon
from the exchanged pre-reggeons is an essential part of our
result.  Whereas the corresponding effective vertices in the case of 
the leading gluons and quarks \cite{BFKL}, \cite{Fad-Sher} 
where known long before the effective action
approach to high-energy scattering was proposed, the effective 
vertices for the next-to-leading exchanges appear here the first time.
To investigate these vertices more closely will be our next task.

The investigations along these lines contribute to a deeper
understanding of the high-energy asymptotics of gauge theories.
There is the hope that some of the interesting structure
uncovered in perturbation theory is significant in general.
On the other hand there are problems arrising from the present
day phenomenology of semi-hard processes to which the next-to-leading
high-energy effective action can be applied.

\vspace*{1cm}
{\Large \bf  Acknowledgements}

\vspace*{.8cm}

L.Sz. would like to acknowledge the warm hospitality extended to him 
during his stay at
University of Leipzig.

%
%

\newpage


\begin{figure}[htb]
\begin{center}
\epsfig{file=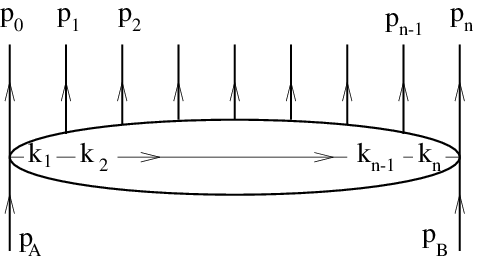,width=8cm}
\end{center}
\vspace*{0.5cm}
\caption{
The inelastic process in the multi-Regge kinematics
}
\end{figure}


\begin{figure}[htb]
\begin{center}
\epsfig{file=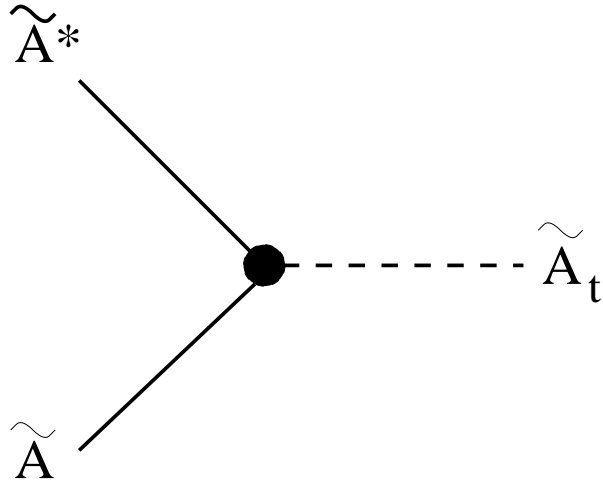,width=8cm}
\end{center}
\vspace*{0.5cm}
\caption{
The graphical illustration of vertices entering Eqs. (2.10) and (2.12).
}
\end{figure}



\begin{figure}[htb]
\begin{center}
\epsfig{file=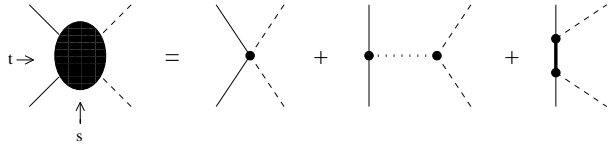,width=8cm}
\end{center}
\vspace*{0.5cm}
\caption{
The graphical illustration of Eq. (4.1) for the complete quartic terms.
Different line forms represent different modes: full line - scattered modes,
dotted line - exchange modes, bold line - heavy modes, dashed line - the sum of
all modes.
}
\end{figure}


\begin{figure}[htb]
\begin{center}
\epsfig{file=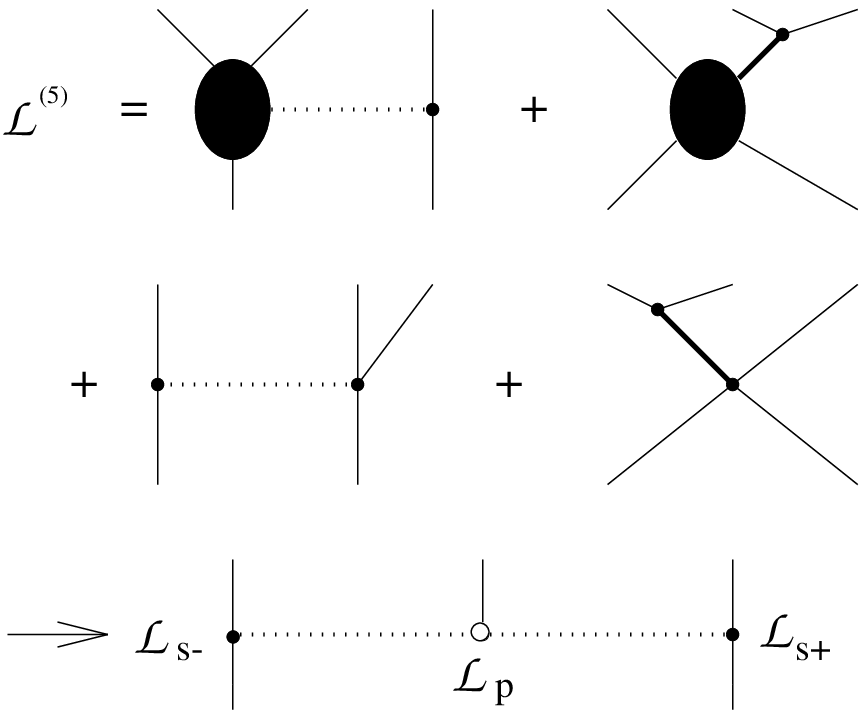,width=8cm}
\end{center}
\vspace*{0.5cm}
\caption{
The graphical illustration of the terms describing inelastic scattering
Eq. (5.2) and
of the factorization leading to the production vertices Eq. (5.1).
}
\end{figure}

\end{document}